 \documentclass[preprint]{aastex}

\shorttitle{Crab Pulsar}
\shortauthors{Crusius-W\"atzel, Kunzl and Lesch}

\begin{document}

\title{
SYNCHROTRON MODEL FOR THE INFRARED, OPTICAL AND X-RAY EMISSION OF THE CRAB PULSAR}

\author{Andr\'e R. Crusius-W\"atzel, Thomas Kunzl and Harald Lesch}
\affil{Institut f\"ur Astronomie und Astrophysik der Universit\"at
M\"unchen\\ Scheinerstr. 1, 81679 M\"unchen, Germany}

\begin{abstract}
We develop a model for the infrared, optical and soft X-ray
emission of the Crab
pulsar in terms of anisotropic synchrotron emission by relativistic particles in an
outer gap scenario with a single energy distribution $N(\gamma)\propto \gamma^{-2}$.
It is shown that such a distribution is naturally produced in an efficient
pair cascade and that the energy of the primary particles are limited
by synchrotron radiation to $\gamma\sim 10^7$.
It is further shown that this synchrotron model is able to reproduce
the spectral shape between the infrared and soft X-rays and also
the corresponding luminosities.
In particular, the long standing problem of the rapid spectral decline towards
infrared frequencies is understandable as emission at very small
pitch angles from low energy particles with $\gamma\sim 10^2$.

Finally, we show that the scaling of our synchrotron model explains
the observed correlation between the X-ray luminosity and the spin-down
luminosity of the neutron star $L_{\rm x} \sim 10^{-3} L_{\rm sd}$
found by \citet{bec97}.

\end{abstract}

\keywords{Stars: Pulsars: Individual (PSR 0531+21) --- Radiation Mechanisms:
Nonthermal (Synchrotron Radiation) --- Stars: X-Rays}

\section{Introduction}

The Crab pulsar has a continuous spectrum from the optical to X-rays and
$\gamma$-rays with different power-laws \citep{lyn90}.
The spectral index $\alpha$, defined as $I_{\nu}\propto
\nu^{-\alpha}$, varies from $\alpha=1.1$ in the $\gamma$-region via
$0.7$ in the hard X-ray region to $0.5$ at soft X-rays \citep{too77} and
zero at optical frequencies. In the far infrared
region the spectrum is inverted $\alpha=-2$ and cuts off sharply towards
lower frequencies.
This behaviour is
not accompanied by dramatic pulse profile changes, as one would expect from
saturation or self-absorption effects. Self-absorption should first
influence the peak intensity. Although \citet{pen82} claims to have
measured about 20\%  peak-depression in the near infrared,
the situation is ambiguous.
Absorption outside the
pulsar environment would not influence the pulse form, but the nature of
such an absorber is not clear. The conventionally used optically thin
synchrotron radiation theory yields rising spectra if the particle spectrum
has a low-energy cutoff, with a spectral index of $-1/3$,  if the angle
integrated single particle emissivity is used and $-2/3$,  if the angular
dependence is retained, e.g., when the emission is spread over angles
becoming comparable to the pulsar emission cone angle. The infrared spectrum
measured by \citet{mid83} therefore is still awaiting an explanation.
It is one of the purposes of this paper to show that the theory of optically thin
synchrotron radiation at very small pitch angles gives a possible solution
to this problem. \citet{epst73} applied small pitch angle synchrotron radiation
to the Crab optical pulsations, but the detailed spectral features were
only calculated for AGN.

\section{Outer Gap Scenario and Particle Energy Spectrum}

We suggest that the infrared, optical and ultraviolett emission (and
also the X-ray and $\gamma$-ray emission) can be explained in terms of
an outer magnetospheric gap model \citep{che86}. Models for the Gamma ray
emission from outer gaps have been published by \citet{ry95} and
\citet{cr94}. This idea is also supported by
the different widths of pulse profiles in the radio (inner magnetosphere) and
the higher frequency range. The magnetic field strength close to the velocity
of light cylinder of Crab pulsar is $B\approx 10^6 {\rm G}$.

In such a scenario  particles are accelerated to high Lorentz factors
$\gamma_p\sim 10^7$ by large potential drops along the magnetic field lines,
usually thought to be limited by curvature radiation. The electric field strength
in the gap of a pulsar rotating with angular velocity $\Omega=2\pi/P$ has been
calculated by \citet{che86} to be
\begin{equation}
E_0={\Omega B a^2\over R_{\rm c} c}=6.3\times 10^3\left({a\over 10^7{\rm cm}}\right)^2
\left({R_{\rm c}\over 10^8{\rm cm}}\right)^{-1}\left({P\over 33{\rm ms}}\right)^{-1}
\left({B\over 10^6{\rm G}}\right)\ {\rm statvolt/cm}\ ,
\end{equation}
where $R_c$ is the curvature radius of the field lines in the gap region and $a$
is the dimension of the gap perpendicular to the magnetic field $B$.
The rate at which particles gain energy in such an electric field is
\begin{equation}
P_{\rm gain}=ceE_0\ ,
\end{equation}
and the energy loss rate due to curvature radiation is
\begin{equation}
P_{\rm cr}=\frac{2}{3}\frac{e^2c}{R_{\rm c}^2}\gamma^4=4.6\times 10^{-9}R_{\rm c}^{-2}
\gamma^4\ {\rm erg\,s^{-1}}\ .
\end{equation}

We now consider a competing process, which can limit the acceleration of particles
to high energies, namely synchrotron radiation. The energy loss rate of a single
particle is then given as
\begin{equation}
P_{\rm syn}=1.6\times 10^{-15}\gamma^2 B^2\sin^2\Psi\ {\rm erg\,s^{-1}}\ ,
\end{equation}
which is larger than the curvature radiation losses if the pitch angle is
\begin{equation}
\Psi\geqslant 1.7\times 10^{-4}\left({\gamma\over 10^7}\right)
\left({B\over 10^6{\rm G}}\right)^{-1}\left({R_{\rm c}\over 10^8{\rm cm}}\right)^{-1}
\ ,
\end{equation}
where $\sin\Psi\approx\Psi$ has been used.

Since we are interested in the acceleration process close to the light cylinder radius
$R_{\rm lc}$, we now calculate the accelerating electric field there.
Let $r$ and $\theta$ be the radial distance from the neutron star center and the
angle between the magnetic moment and the radius vector to a point in the
magnetosphere, respectively. The angle between the magnetic moment and the rotation
axis is called $\alpha$. Let further be $x=r/R_{\rm ns}$,
where $R_{\rm ns}\approx 10^6$~cm is
the radius of the neutron star. The radius of the light cylinder $R_{\rm lc}$ is given by
$\Omega R_{\rm lc}=c$, so that
\begin{equation}
x_{\rm lc}={cP\over 2\pi R_{\rm ns}}\ .
\end{equation}
In a purely dipolar magnetic field the relation
\begin{equation}
{\sin^2\theta\over r}={\sin^2\theta_0\over R_{\rm ns}}
\end{equation}
holds for every field line,
where $\theta_0$ defines its footpoint at the neutron star surface.
The curvature radius of a magnetic field line is known to be
\begin{equation}
R_{\rm c}={ R_{\rm ns}\over\sin\theta_0}x^{1/2}\ .
\end{equation}
In an aligned rotator the last closed field line would be characterized by $\theta=\pi/2$
at the light cylinder radius,
whereas in the non-aligned case it is given by $\theta=\pi/2-\alpha$. It is then
easily verified with eq.~(7), that
\begin{equation}
\sin\theta_0=x_{\rm lc}^{-1/2}\cos\alpha\ .
\end{equation}
We therefore find for the curvature radius at the light cylinder
\begin{equation}
R_{\rm c}={ R_{\rm ns}x_{\rm lc}\over\cos\alpha}={ R_{\rm lc}\over\cos\alpha}\ .
\end{equation}
The accelerating electric field near the light cylinder is then
\begin{equation}
E_0={a^2\over R_{\rm lc}^2}\cos\alpha\thinspace B=\delta^2\cos\alpha\thinspace B\ ,
\end{equation}
which defines $\delta$ as the ratio of the gap thickness to the light cylinder radius,
in the following assumed to be of order 0.1.

One can now estimate the equilibrium Lorentz factor for which energy gains and
synchrotron losses are equal, as
\begin{equation}
\gamma_{\rm p}=9.5\times 10^6\cos^{1/2}\alpha\left({\delta\over 0.1}\right)
\left({B\over 10^6{\rm G}}\right)^{-1/2}\left({\Psi\over 10^{-3}}\right)^{-1}\ .
\end{equation}
These particles are called primary in the following. By inverse Compton
scattering they can produce $\gamma$-rays with energies up to 5 TeV.
We use a pitch angle of order $\Psi\sim 10^{-3}$ which in the following sections
is shown to be required to reproduce the infrared and optical spectrum, as well
as the luminosity in this band.

Due to interactions of these $\gamma$-rays with photons a pair creation cascade
is ignited, which can produce several generations of positrons and electrons
of equal amount and a certain multiplicity $M$ compared to the
Goldreich-Julian density. The initial momentum of the created pairs will be
parallel to that of the primary particle, i.e. the pairs will also have a pitch angle
$\Psi\simeq 10^{-3}$. In typical cascade models the pair
density exceeds the density of primaries by a factor of $10^3\dots10^5$
\citep{m95}. Numerical simulations of such a cascade showed
that a power-law energy distribution $N(\gamma)=N_0\gamma^{-s}$ will be
produced, with a typical index of $s\approx 2$. The first cascade considerations
were performed by \citet{dh82} for an inner gap
scenario but if the cascade is possible in the outer gap (which is commonly
assumed) the energy spectrum of the secondaries should not be too different.
The spectrum cuts off below $\gamma\lesssim 10^2$ due to the energy theshold
of the interacting photons to produce a pair.

This particle energy spectrum can also be found by the following simple argument.
Consider a primary particle with energy $\gamma_{\rm p} m_e c^2$ cascading into
$M(\gamma)$ particles, each with an energy $\gamma m_ec^2$. If the cascade
is very efficient, energy conservation leads to
\begin{equation}
M(\gamma)={\gamma_{\rm p}\over\gamma}\sim{10^7\over\gamma}\ .
\end{equation}
Since $M(\gamma)$ is the number of particles around Lorentz factor $\gamma$,
the differential number density, or energy spectrum is then
\begin{equation}
N(\gamma)\propto {M(\gamma)\over \gamma}\propto {1\over\gamma^2}\ .
\end{equation}

The initial momentum of the primary particle (parallel to the local
magnetic field) defines also the momentum of the $\gamma$-rays emitted by
inverse Compton scattering with ambient photons. The $\gamma$-rays eventually
leave the gap because they are not influenced by the magnetic field
and proceed to created pairs by photon-photon interactions. The subsequent
generations of pairs are then no longer affected by the electric field of the gap.
So it seems reasonable to assume that via creation all particles have
the same pitch angle with a following evolution of the pitch angle distribution
due to motion along the magnetic field lines and radiative losses.
Of course the emisssion
of photons will influence the particle energy spectrum, giving a steepening
after a loss time scale $t_{\rm syn}$, affecting first the high energy part.

\section{Synchrotron Emission at Very Small Pitch Angles}

Since the relativistic particles of the pair cascade are created almost
in the direction of the magnetic field lines one has to be careful to
use the correct validity range of the approximations made in
synchrotron radiation theory.
In \cite{eps73} the
synchrotron emission process by particles with very small pitch angles is
discussed in detail.
The emissivity at very low pitch angles ($\Psi\ll 1/\gamma$) of a single
particle is given by
\begin{equation}
\epsilon_{\nu}(\theta,\gamma)
={\pi e^2\gamma\Psi^2\nu^3\over{\nu_Bc}}\left[1-{\nu\over{\gamma\nu_B}}+
{\nu^2\over{2\gamma^2\nu_B^2}}\right]
\delta\left(\nu-{2\gamma\nu_B\over{1+\theta^2\gamma^2}}\right)\ ,
\end{equation}
where $\nu_B=eB/2\pi mc=2.80\times 10^{12}B_6\thinspace {\rm Hz}$
is the nonrelativistic gyro frequency and $\theta$ is the
angle between the magnetic field $B$ ($B_6$: $B$ in units of $10^6 {\rm G}$)
and the direction of emission. Since we restrict our discussion to small
pitch and emission angles we set $\sin\Psi\approx\Psi$ and $\sin\theta\approx\theta$.
This formula has to be applied when angles $\theta\la 1/\gamma$ are
resolved in the observations, i.e. in the case of a pulsar, when the pulse
width $\Delta\phi$ becomes comparable to the emission cone angle of the
particles, $1/\gamma\ga 2\pi\Delta\phi$.
Since the maximum of the emission is in the forward direction ($\theta=0$)
the emission is dominated by the field lines that point towards the observer
at each phase of the pulse. The degree of circular and linear polarization is
given by
\begin{equation}
\rho_c=\mu\thinspace{1-\theta^4\gamma^4\over{1+\theta^4\gamma^4}}
\end{equation}
and
\begin{equation}
\rho_l={2\theta^2\gamma^2\over{1+\theta^4\gamma^4}}
\end{equation}
where $\mu=\pm 1$ is the sign of the charge.

When the typical angle of emission $\theta$ is small compared to
the pulse width it is useful to integrate the emissivity
over all angles
to get the total emission spectrum of a single particle
\begin{equation}
\varepsilon_{\nu}(\gamma)=2\pi\int\epsilon_{\nu}\sin\theta\ d\theta
={2\pi^2e^2\Psi^2\over{c}}\nu\left[1-{\nu\over{\gamma\nu_B}}+
{\nu^2\over{2\gamma^2\nu_B^2}}\right]
\end{equation}
up to $\nu\leq2\gamma\nu_B$. The energy loss rate of a single particle
is obtained by integrating
the total emission over all frequencies
\begin{equation}
L_{\rm p}=-{dE\over dt}=\int_0^{\infty}\varepsilon_{\nu}\ d\nu
={8\pi^2e^2\nu_B^2\gamma^2\Psi^2\over{3c}}\ ,
\end{equation}
which is also the well known result at large pitch angles, used in eq.~(4).

We now calculate the emission from a power law distribution
of relativistic particles, $N(\gamma)=N_0\gamma^{-s}$,
in the direction along the magnetic field $\theta=0$
according to
\begin{eqnarray}
I_{\nu}\propto\int\epsilon_{\nu}(0,\gamma) N(\gamma)\thinspace d\gamma
&=&{\pi e^2\Psi^2N_0\nu^3\over{2\nu_B^2c}}\int\delta\left(\gamma-{\nu\over2\nu_B}
\right)\gamma^{1-s}\thinspace d\gamma\\
&=&{4N_0\pi e^2\Psi^2\nu_B\over c}\left({\nu\over2\nu_B}\right)^{4-s}\ .
\end{eqnarray}
This result shows that the spectrum rises very steeply, in fact
$I_{\nu}\propto\nu^2$ for $s=2$. Therefore the sharp rise in the spectrum of
the Crab pulsar at infrared frequencies can be explained as synchrotron
emission at very small pitch angles. The Lorentz factors in the case of the
Crab then have to be of the order of $\gamma\la 1/\Delta\phi\sim 10^2$ so that
this description applies. This is consistent with the requirement
that $2\gamma\nu_B$ lies in the infrared part of the spectrum.

In case of somewhat higher Lorentz factors the single particle
beam angle becomes smaller than the phase angle of the pulse and the total
(or angle averaged) power spectrum has to be applied.
To find the spectrum from a power-law
distribution the monochromatic approximation
\begin{equation}
\varepsilon_{\nu}\approx -{dE\over dt}\thinspace\delta(\nu-2\gamma\nu_B)
\end{equation}
is used for simplicity. This is a good approximation
since the emission is sharply peaked at $2\gamma\nu_B$ and one may just put
all the emission at this frequency. As a result we then find
\begin{equation}
I_{\nu}\propto\int\varepsilon_{\nu} N(\gamma)\ d\gamma
={4\pi^2e^2N_0\nu_B^2\Psi^2\over{3c}}\left({\nu\over2\nu_B}\right)^{2-s}\ .
\end{equation}
In the case of $s=2$ this gives the observed flat, $\alpha\approx 0$,
optical/UV spectrum of the Crab pulsar, with $\gamma\sim 10^2-10^3$.
Since the flat spectrum continues up to the far-ultraviolett
\citep{lun99} the pitch angle involved has to be $\Psi\sim 10^{-3}$ so that
the emission is still at very low pitch angles ($\gamma\Psi< 1$).

\section{Crab Pulsar Spectrum from the Infrared to X-rays}

We consider now the injection of a power law distribution of a relativistic
electron-positron plasma $N(\gamma)=N_0\gamma^{-2}$, which is
produced in the cascading process, along the magnetic field lines.
We interpret the infrared and optical emission of the crab pulsar
with its fast rise
towards the optical as synchrotron emission from relativistic particles
with very small pitch angles ($\gamma\ll 1/\Psi$).
As shown in the previous section we expect a spectral index of $\alpha=-2$
in the infrared part and $\alpha=0$ in the optical part, if the pitch angle
is of order $\Psi\sim 10^{-3}$ for Lorentz factors $\sim 10^2\,-\,10^3$.

Although the circular polarization will be at
its maximum when looking along the field lines, in the cascade process
an equal amount of
electrons and positrons will be produced, giving a net zero circular
polarization. On the other hand the linear polarization becomes maximal for
$\theta=1/\gamma$ for both species and is zero for $\theta=0$. If the
emission cone of the pulsar is not cut right through the center by the line
of sight but slightly offset, then one expects a polarization angle swing.

The soft X-ray emission in this model
comes from the particles with larger pitch angles ($\Psi\gg 1/\gamma$),
radiating at a typical frequency
\begin{equation}
\nu_c={3\over 2}\nu_B\Psi\gamma^2
\end{equation}
The spectrum in this large pitch angle limit is given by
\begin{equation}
I_{\nu}\propto \nu^{-(s-1)/2}
\end{equation}
which yields the
observed $I_{\nu}\propto\nu^{-0.5}$ at soft X-rays for $s=2$. Therefore a
single power-law distribution of relativistic electrons and positrons can
produce the complete spectrum from infrared to X-ray frequencies.

\section{Simple Luminosity Estimates for the Optical and X-ray Emission}

We now estimate the energy radiated by the pulsar in the optical and in the
X-ray band simply as the number of particles times the energy radiated by
one particle. Thus we have
\begin{equation}
L=Mn_{\rm GJ}VP_{\rm syn}
\end{equation}
with
\begin{equation}
P_{\rm syn}=1.6\times10^{-15}\Psi^2\gamma^2B^2\ {\rm erg\thinspace s^{-1}}
\end{equation}
being the energy radiated by a single particle. $M$ is the multiplicity of the
relativistic particle density wihin the volume $V$
as compared to the Goldreich-Julian density
\begin{equation}
n_{\rm GJ}=6.9\times 10^{-2} B\thinspace P^{-1}\ {\rm cm^{-3}}\ ,
\end{equation}
where $P$ is the period of the pulsar in seconds, $x=R/R_{\rm ns}$ is the
distance in units of the neutron star radius. The magnetic dipole field
falls off as $B=B_0/x^3$, with $B_0$ being the field strength at the surface
of the neutron star.
Since we assume the radiation to
occur close to the light cylinder we set
\begin{equation}
x=x_{\rm lc}=4774\thinspace P\ .
\end{equation}
The volume is estimated as a spherical shell within a fraction $f$ of
the light cylinder radius
\begin{equation}
V=4\pi R_{\rm lc}^2\cdot fR_{\rm lc}=1.4\times 10^{30}fP^3\ {\rm cm^{3}}\ .
\end{equation}
With $M=\gamma_{\rm p}/\gamma$ we then find that
\begin{equation}
L=1.5\times 10^{14}\gamma_{\rm p}\gamma f\Psi^2B^3P^2\ {\rm erg\thinspace s^{-1}}\ .
\end{equation}
The frequency of the optical emission (at small pitch angles) is given by
\begin{equation}
\nu_{\rm opt}=2.8\times 10^6\gamma_{\rm opt} B\ {\rm Hz}\ ,
\end {equation}
so that, by eliminating $\gamma$, we get
\begin{eqnarray}
L_{\rm opt}=4\times 10^{31}\cos^{1/2}\alpha
\left({\delta\over 0.1}\right)
\left({f\over 0.1}\right)
\left({\Psi_{\rm p}\over 10^{-3}}\right)^{-1}
\left({\Psi_{\rm opt}\over 10^{-3}}\right)^2
\left({\nu_{\rm opt}\over 10^{15}{\rm Hz}}\right)\nonumber\\
\times
\left({B_{0}\over 10^{12.5}{\rm G}}\right)^{3/2}
\left({P\over 33{\rm ms}}\right)^{-5/2}\ {\rm erg\thinspace s^{-1}}\ .
\end{eqnarray}
The parameter values chosen for the Crab pulsar are consistent with the
considerations of the spectral shape and the cascading process,
i.e. the particles with $\gamma\sim 10^2-10^3$ have pitch angles
$\Psi_{\rm opt}\approx 10^{-3}$. This crude estimate gives the right order of
magnitude of the Crab pulsars optical luminosity.

In the X-ray regime where the large pitch angle approximation for the
frequency applies
\begin{equation}
\nu_{\rm x}=2.8\times 10^6\Psi_{\rm x}\gamma_{\rm x}^2B\ {\rm Hz}
\end{equation}
we then find
\begin{eqnarray}
L_{\rm x}=6\times 10^{35}\cos^{1/2}\alpha
\left({\delta\over 0.1}\right)
\left({f\over 0.1}\right)
\left({\Psi_{\rm p}\over 10^{-3}}\right)^{-1}
\left({\Psi_{\rm x}\over 10^{-1}}\right)^{3/2}
\left({\nu_{\rm x}\over 10^{17}{\rm Hz}}\right)^{1/2}\nonumber\\
\times
\left({B_{0}\over 10^{12.5}{\rm G}}\right)^{2}
\left({P\over 33{\rm ms}}\right)^{-4}\ {\rm erg\thinspace s^{-1}}\ .
\end{eqnarray}
In order to get the right X-ray luminosity a pitch angle of order
$\Psi_{\rm x}\sim 0.1$ is required. To emit at soft X-ray frequencies
again the Lorentz factors have to be in the range $\gamma\sim 10^2-10^3$.

This result can now be compared with the spin-down luminosity of the Crab pulsar
\begin{equation}
L_{\rm sd}=3.2\times 10^{38}
\left({B_{0}\over 10^{12.5}{\rm G}}\right)^{2}
\left({P\over 33{\rm ms}}\right)^{-4}
\ {\rm erg\thinspace s^{-1}}\ .
\end{equation}
The luminosity radiated in X-rays by the synchrotron mechanism shows the
same dependence on the magnetic field strength and the period as the
rotational energy loss of the pulsar. We can therefore write
\begin{equation}
L_{\rm x}\approx 2\times 10^{-3}\cos^{1/2}\alpha
\left({\delta\over 0.1}\right)
\left({f\over 0.1}\right)
\left({\Psi_{\rm p}\over 10^{-3}}\right)^{-1}
\left({\Psi_{\rm x}\over 10^{-1}}\right)^{3/2}
\left({\nu_{\rm x}\over 10^{17}{\rm Hz}}\right)^{1/2}
\thinspace L_{\rm sd}\ .
\end{equation}
In order to get the right X-ray luminosity a pitch angle of order
$\Psi_{\rm x}\sim 0.1$ is required. To emit at soft X-ray frequencies
again the Lorentz factors have to be in the range $\gamma\sim 10^2-10^3$.

Since this relation is now independent of $B_0$ and $P$ it should apply to
any pulsar, not only the Crab pulsar.
The observations of X-ray selected pulsars by \citet{bec97} have revealed
such a dependence in the form $L_{\rm x}\sim 10^{-3}L_{\rm sd}$.
The theoretically found proportionality
between the X-ray luminosity and the spin-down luminosity strongly supports
our assumption, that synchrotron emission is the dominant radiation process.
On the basis of the model presented here, this observation implies that
the parameter combination $\cos^{1/2}\alpha\thinspace \delta f\Psi_{\rm p}^{-1}
\Psi_{\rm x}^{3/2}$ has about the same value for every pulsar in their list.

The luminosity estimates also contain the correct spectral dependencies, because
$I_{\nu}\propto L_{\rm opt}/\nu\propto const.$ for the optical and
$I_{\nu}\propto L_{\rm x}/\nu\propto \nu^{-1/2}$ for the X-ray emission.

\section{Discussion and Conclusions}

The high frequency emission of the Crab pulsar is a challenge
for the theoretical description.
Earlier investigations put the location of the infrared and optical pulses
close to the light cylinder radius \citep{smi88}
and led us to the idea that these distant
regions of the magnetosphere may be responsible for the emission up to the
X-ray and gamma ray range. First we were able to show that
within an outer gap description and
a standard cascade model we expect a particle energy distribution
$N(\gamma)\propto \gamma^{-2}$, being the basis for the investigations of
the radiative properties of the Crab pulsar. The energy of the primary
particles accelerated in the electric field of the gap is shown to be
limited by synchrotron radiation and not curvature radiation
to $\sim 10^7m_{\rm e}c^2$. This holds for pitch angles $\Psi\gtrsim 10^{-4}$.

A detailed inspection of an anisotropic synchrotron model led to the result
that the emission of particles with small pitch angles with this single
energy distribution reproduces the key properties of the
spectrum and luminosity from the infrared up to soft X-rays.
This model explains all relevant features, including the
long-standing problem of the increase of the infrared spectrum $I_\nu\propto \nu^2$,
the flat spectrum $I_\nu\propto \nu^0$ in the optical and UV range and the
X-ray spectrum $I_\nu \propto \nu^{-0.5}$. The only difference concerns the
pitch angle of the emitting particles and the (energy
dependent) multiplicity $M=M(\gamma)$.
The infrared/optical/UV part of the spectrum is emitted by particles with
Lorentz factors $\gamma\sim 10^2-10^3$ and
pitch angle $\Psi\sim 10^{-3}$, corresponding to the pitch angle of the
primary particles accelerated in the electric field of the gap.
The X-rays are radiated by particles in the same energy range but
with a pitch angle $\Psi\sim 0.1$. It is not clear what the fundamental
physical scenario is, which gives the required values for the pitch angle.

The X-ray emission from particles with
the smaller pitch angles, involving higher energy particles, which are smaller
in number, would not be seen due to the smaller power radiated.
It is dominated by particles
with the larger pitch angle. On the other hand for pulsars with smaller
magnetic field strength at the light cylinder, the emission at
larger pitch angles might be seen in the optical band. In that case
the $\nu^{-1/2}$-spectrum is also expected there, then accompanied by a
luminosity relation similar to that for the X-rays.

From simple luminosity arguments we find that this model is able to yield
the desired energy output in the optical as well as in the X-ray part of the
spectrum.
We also find that the observed proportionality of the X-ray flux and the spin-down
luminosity is a consequence of the synchrotron radiation model for any
pulsar that developes an efficient pair cascade.

\acknowledgements{\sl Acknowledgements:}
This work was supported by the Deutsche Forschungsgemeinschaft with the grant
LE 1039/6 (C.-W.).

\end{document}